# Mass Additivity and A Priori Entailment

Kelvin J. McQueen

*The principle of mass additivity states that the mass of a composite object is the sum of the masses of its elementary components. Mass additivity is true in Newtonian mechanics but false in special relativity. Physicists have explained why mass additivity is true in Newtonian mechanics by reducing it to Newton's microphysical laws. This reductive explanation does not fit well with deducibility theories of reductive explanation such as the modern Nagelian theory of reduction, and the a priori entailment theory of reduction that is prominent in the philosophy of mind. Nonetheless, I argue that a reconstruction of the explanation that incorporates distinctively philosophical concepts in fact fits both theories. I discuss the implications of this result for both theories and for the reductive explanation of consciousness.*

## 1. Mass Additivity

The principle of mass additivity states that the mass of a composite object is the sum of the masses of its elementary components. Mass additivity is true in Newtonian mechanics and physicists have explained mass additivity in terms of Newton's microphysical laws. However, mass additivity is false in special relativity.[1] Accordingly, there are interesting philosophical questions about the status of mass additivity: is mass additivity a fundamental axiom of Newtonian mechanics or is it derivative? Since physicists have reductively explained mass additivity in terms of Newton's microphysical laws, it would seem the principle is derivative. However, the explanation as it stands is problematic, or so I shall argue.

This article has two primary goals. The first concerns the explanation of mass additivity, the second concerns the explanation's significance to philosophy. Regarding the first goal, the standard textbook explanation is presented as a deduction of mass additivity from principles of Newtonian microphysics (sec. 2). However, the deduction is only valid for a restricted class of Newtonian bodies, and it is not immediately clear how the explanation should be generalised (sec. 3). Here I attempt to solve this problem. Thus, the first goal is to offer what may be the most complete explanation of mass additivity in Newtonian mechanics (sec. 4-5).

---

[1] In special relativity the mass $m_c$ of a composite composed of $N$ particles each with energy $E_i$ and momentum $p_i$ is $m_c = \left[\left(\sum_{i=1}^{N} \frac{E_i}{c^2}\right)^2 - \left(\sum_{i=1}^{N} \frac{p_i}{c}\right)^2\right]^{1/2}$. See Okun (1989: 632; 2000: 1271).

The second goal is to relate the analysis to contemporary discussions of reductive explanation in the philosophy of science and the philosophy of mind. The explanation of mass additivity is useful for testing philosophical accounts of explanation because although the details of the example are non-trivial, the relevant science is at the same time very well understood. It is therefore surprising that the explanation has not been considered by philosophers. The problems I raise for the explanation concern the fact that mass additivity has not been deduced, in full generality, from the explanans. So it is an interesting case study for modern deducibility theories of explanation.

Deducibility theories of explanation divide an explanation into an explanans (the explaining statements) and an explanandum (the statement to be explained). The *core deducibility theory* states that a reductive explanation is successful if and only if the explanandum is logically deducible from the explanans in conjunction with analytic definitions of the terms in the explanandum. Different deducibility theories weaken the core theory with different qualifications, while retaining significant emphasis on the role of deduction. I consider two such theories. Firstly, Nagel's (1949, 1961, 1979) resurging theory of reductive explanation, particularly as it is now commonly understood, in terms of the Generalised Nagel-Schaffner (GNS) theory.[2] Secondly, the a priori entailment theory of reduction (AET) that is prominent in the philosophy of mind, and which has been used to argue for the irreducibility of consciousness.[3]

In section 6 I argue that GNS fails to account for the explanation of mass additivity as it has been expressed by physicists. This is despite the fact that the explanation falls into a neglected category of reduction that for Nagel "generates no special logical puzzles" (1961: 339). However, I argue that this need not refute GNS. I offer an explanation of mass additivity that *does* fit the GNS model. It is therefore open to Nagelians to maintain that the explanation offered here is the correct explanation. However, the explanation offered here requires distinctively philosophical concepts. Nagelians therefore cannot deny the significance of the class of explanation to which mass additivity belongs.

In section 7 I explain that AET can account for the scientific explanation of mass additivity provided that the explanation is suggestive of an a priori entailment of the explanandum from the explanans (perhaps together with other lower-level truths). Since the explanation I offer generalises the scientific explanation and demonstrates an a priori entailment of mass additivity from microphysics, I conclude that the example supports AET. However, the analysis also shows how lower-level reducing facts can trump a priori intuitions about the reducibility or otherwise of higher-level properties. I use this to challenge the use of AET to argue for the irreducibility of higher-level properties, such as consciousness, when lower-level facts are not fully understood.

---

[2] The name comes from Dizadji-Bahmani et. al. (2010: sec. 3.1) who contrast this view with Schaffner's (1993, ch. 9) own so-called *generalized reduction-replacement model*. The latter is stronger as it involves specific claims about bridge principles that the former stays neutral on. See also Schaffner (2012).
[3] For AET see Chalmers and Jackson (2001), Gertler (2002), and Chalmers (2012). For the use of AET to argue for the irreducibility of consciousness see Chalmers (1996).

Before we begin, I make two clarifications regarding the relevant notion of 'mass'. Firstly, there are two primary concepts of mass in Newtonian mechanics: *gravitational* mass and *inertial* mass. While inertial mass is (roughly) a measure of an object's disposition to resist changes in motion given a force, gravitational mass is (roughly) a measure of an object's disposition to exert (or experience) an attractive force on other (gravitationally massive) objects (Jammer 2000: 6). Gravitational mass and inertial mass are conceptually distinct, but *a posteriori* equivalent (Pockman 1951). I do not have space to discuss gravitational mass additivity but believe that much of what is said here is applicable. In what follows every use of 'mass' refers to inertial mass.

Finally, because I claim that mass additivity is false in special relativity, some may refer to what I am calling 'mass' as 'rest mass' (or 'proper mass' or 'invariant mass') so as to distinguish it from what they call 'relativistic mass'. Relativistic mass is additive whereas rest mass obeys the relativistic equation for composite mass from above. I join Einstein (1948) and most contemporary physicists in rejecting the notion of relativistic mass[4] and consider rest mass (i.e. *mass*) to be the referent of Newton's term 'mass'. It is in this sense of 'mass' that we have discovered a posteriori that what Newton believed to be additive is in fact non-additive. All of the considerations raised here regarding mass additivity in Newtonian mechanics can equally well be raised for the relativistic equation for composite mass, which physicists have at least partially explained in terms of relativistic microphysics (Gabovich and Gabovich: 2007).

## 2. The Textbook Reductive Explanation of Mass Additivity

According to Norman Feather (1966: 511) the very coherence of Newtonian mechanics depends on a microphysical reduction of mass additivity:

> "Newton was an atomist, and it is clear that he conceived of the forces which he identified as operating between bodies, and whose interrelations he described in the third law, as compounded of the elementary forces acting between the particles (primordial atoms) of which real bodies are constituted. [...] It is clear, then, that the question of the additivity of mass is fundamental for the logical development of the Newtonian scheme: without its consideration, the transition from particles to gross bodies cannot be made convincingly, and the simple mechanics of particles fails to achieve relevance to the real world."

---

[4] "As regards active specialists they answer in perfect unison insofar as their scientific work is concerned: the mass is independent of velocity, it is not additive […] there is no disagreement among researchers on the definition of mass. […] According to modern terminology, both terms, 'relativistic mass' and 'rest mass', are obsolete" (Okun 2000: 1270). Compare Field (1973: 469) and Feyerabend (1962: 80-1).

Some classical physics textbooks provide reductive explanations of mass additivity and the explanations typically take the same general form.[5] The following explanation, taken from Kibble and Berkshire's *Classical Mechanics*, aims to show that "one consequence of our basic laws is the additive nature of mass" (2004: 12). It is clear that Kibble and Berkshire (hereafter: K&B) are offering a reductive explanation. After all, their notion of 'consequence' is not a causal notion—they are not trying to show that a composite's mass is a causal consequence of anything. Hence, they propose to reduce mass additivity to the basic laws.

K&B consider an isolated three body system in light of Newton's second law ($F_i = m_i a_i$) and the superposition principle for interactions ($F_i = \sum_{j=1}^{N} F_{ij}$). The former states that the force on particle $i$ is equivalent to the product of that particle's mass and acceleration. The latter states that the force on particle $i$ is equivalent to the sum of each of the individual forces $F_{ij}$, where $F_{ij}$ stands for the force on particle $i$ due to the particle indexed by $j$ ($j$=1,2,...,$N$). Applying these microphysical laws to the three particles gives:

$$F_{12} + F_{13} = m_1 a_1 \tag{1}$$

$$F_{21} + F_{23} = m_2 a_2 \tag{2}$$

$$F_{31} + F_{32} = m_3 a_3 \tag{3}$$

Now Newton's third law ($F_{ij} = -F_{ji}$), which states that there is a force on particle $i$ due to particle $j$ if and only if there is an equal and opposite force on $j$ due to $i$, is introduced. The third law shows that if we add (1)-(3) then the terms on the left cancel in pairs:

$$m_1 a_1 + m_2 a_2 + m_3 a_3 = 0 \tag{4}$$

This gives an equation for $m_1 a_1$:

$$m_1 a_1 = -m_2 a_2 - m_3 a_3 \tag{5}$$

Now comes a crucial auxiliary assumption regarding component accelerations, before the final conclusion is drawn:

> "if we suppose that the force between the second and third is such that they are rigidly bound together to form a composite body, their accelerations must be equal: $a_2 = a_3$. In that case, we get
>
> $$m_1 a_1 = -(m_2 + m_3) a_2 \tag{6}$$
>
> which shows that the mass of the composite body is just $m_{23} = m_2 + m_3$" (2004: 12).

---

[5] Examples are Kibble and Berkshire (2004: 11-12), and Lindsay (1961: 19-21). Feather (1965) is an important attempt to generalise these explanations, see section 4 below.

Why do K&B think that (6) *shows* that the composite's mass is additive? The basic idea must be this: particles 2 and 3 compose a composite - call it 'C'. In general, when a composite's parts each have acceleration $a$ then the composite has acceleration $a$. So the right hand side of (6) contains a term for the composite's acceleration yielding: $m_1 a_1 = -(m_2 + m_3) a_C$. Now consider the left hand side of (6) and the fact that $m_1 a_1 = F_1$. If we could deduce $F_1 = -F_C$ from the microphysical explanans then we would have a crucial expression relating *the force on the composite* and *the composite acceleration* to *the sum of the masses of its parts*:

$$F_c = (m_2 + m_3) a_c \tag{7}$$

Thus, granting that $F_1 = -F_C$, K&B deduce from microphysics both the force and acceleration of the composite and then use the second law to solve for its mass, the value of which is suggested by (6), and is made clear by (7).

### 3. Problems for the Explanation of Mass Additivity

I isolate three problems for the explanation. The Many-Particle Problem concerns the fact that K&B only prove the additive mass of a *two*-particle composite. The Force Additivity Problem concerns the assumption $F_1 = -F_C$. The Composite Acceleration Problem concerns the assumption that components have identical accelerations. All three problems concern the apparent inability to deduce mass additivity in full generality from the microphysical explanans.

The Many-Particle Problem arises because from the explanans one cannot deduce any facts about composites composed of more than three particles, let alone their masses. This can be solved by applying K&B's procedure to $N+1$ particles. In this more general scenario particle 1 interacts with $N$ particles so that we can deduce the mass of the $N$-particle composite. In particular, equation (5) can be written more generally:

$$m_1 a_1 = -\sum_{i=2}^{N+1} a_i m_i \tag{8}$$

Assuming the Force Additivity Problem can be met (i.e. that we can deduce $F_1 = -F_c$ from microphysics) we deduce:

$$F_c = \sum_{i=2}^{N+1} a_i m_i \tag{9}$$

We are yet to confront the Composite Acceleration Problem and so the accelerations of the $N$ component particles are identical. This enables us to bring an acceleration term out of the scope of the sum:

$$F_c = a_2 \sum_{i=2}^{N+1} m_i \qquad (10)$$

Since component accelerations are identical, composite acceleration $a_c$ is identical to $a_2$:

$$F_c = a_c \sum_{i=2}^{N+1} m_i \qquad (11)$$

Given (11) we know that the sum of the component masses is the coefficient relating composite force to composite acceleration. So (11) entails that C's changes in motion given a force is proportional to the sum of the component masses. Thus C's disposition to resist changes in motion given applied forces *just is* the sum of the component masses. So for any number of particles, mass additivity is deducible from Newtonian microphysics together with $F_1 = -F_C$ and the assumption of identical component accelerations.

The remaining two problems are:

**Force Additivity Problem**: It is not clear how to deduce the force exerted by the composite or the force exerted on the composite, from the microphysical explanans. In particular it is not clear how to deduce $F_1 = -F_C$.

**Composite Acceleration Problem**: It is not clear how to deduce the masses of composites whose components do not have identical accelerations, from the microphysical explanans.

Although the philosophical discussion to follow requires a solution to the Force Additivity problem, the discussion will not depend on the specific details of the solution. I therefore deduce $F_1 = -F_C$ from the microphysics in the appendix. I now consider two possible solutions to the Composite Acceleration Problem, the centre of mass solution and the naturalness solution. I argue that only the latter is consistent with the core deducibility theory of reduction.

## 4. The Centre of Mass Solution to the Composite Acceleration Problem

The Composite Acceleration problem is addressed by Norman Feather who tries to "deduce a result valid without approximation" (1966: 511). Feather defines composite position as a particular spatial point, the composite's centre of mass $x_{cm}$. Where $x_i$ is the position of component $i$, the centre of mass of our two-particle composite is:

$$x_{cm} = \frac{m_2 x_2 + m_3 x_3}{m_2 + m_3} \tag{12}$$

The composite acceleration $a_c$ is then defined as the acceleration of the centre of mass $a_{cm}$:

$$a_{cm} = \frac{d^2}{dt^2} \frac{m_2 x_2 + m_3 x_3}{m_2 + m_3} \tag{13}$$

Composite position and acceleration are defined this way despite the components being spatially scattered with distinct accelerations. K&B's procedure is then applied to $a_{cm}$ to deduce composite mass. In particular, if $a_2 \neq a_3$, then we replace (7) with:

$$F_c = (m_2 + m_3) a_{cm} \tag{14}$$

Crucially, (14) holds no matter how distinct the component accelerations are. So it looks like we can deduce mass additivity in full generality, provided that $x_c = x_{cm}$.

But Feather does not argue that $x_c = x_{cm}$. So this is not a deduction of mass additivity from microphysics alone. It is a deduction of mass additivity from microphysics *plus* the stipulation that $x_c = x_{cm}$. Assuming the core deducibility theory of reduction there are two problems.

Firstly, $x_c = x_{cm}$ is not deducible from microphysics. It is impossible for a composite to be located at a single point, if the composite's parts are located at distinct points. Such a composite is patently multiply located at a set of distinct points. So one cannot deduce from microphysics the statement that the composite is located at a point (*any* point). On the contrary it is the negation of this statement that is deducible.

The second problem is that the deduction involves circularity. This is because the centre of mass formula in (12) is derived from facts about composites, including composite mass (as explained below). So the property we are trying to derive from microphysics must be presupposed in order to derive (12). Since the goal is to derive everything from the reduction base - which Feather takes to be purely microphysical given his atomism - Feather's explanation does not achieve this goal.

(12) is meaningful in Newtonian mechanics because Newton's laws applied to it can be used to calculate composite motion. It is this calculation that necessarily invokes composite mass. As Marc Lange (2002: 234) writes:

> "Let us see why it is the case in classical physics that Newton's second law [...] "scales up" in that it governs the motions not only of the elementary bodies, but also of the centre of mass of a system of those bodies. [...] For Newton's second law to govern this system's motion would be for the force exerted on the system (namely, the sum of the forces exerted on the system's constituents [...] to equal the system's mass (m2 + m3; in classical physics, mass is additive) multiplied by the acceleration $a$ of its centre of mass."

If we do not yet know composite mass but only know composite force, then different choices for "the centre" will yield different results for composite mass. For example, if we chose the following point for the position $x_c$ of the composite:

$$x_c = \frac{\sum_i m_i x_i}{\sqrt{\sum_i m_i^2}} \qquad (15)$$

then deducing composite mass $m_c$ by this method (i.e. by dividing composite force by the acceleration of (15)) would force us to deny mass additivity and conclude instead that:

$$m_c = \sqrt{\sum_i m_i^2} \qquad (16)$$

So the very significance of (12) presupposes mass additivity. It is therefore no surprise that in special relativity, where mass additivity fails, a different centre of mass definition is required: "a system's centre of mass is again a weighted average of its constituents' positions but relativity weights them by their energies" (Lange 2002: 235).

The definition in (15) is unnatural: the composite position depends on the choice of origin and can be found well outside the space between the parts.[6] So perhaps one could rule out (15) *a priori* and then argue that if we treat composite position as a point so as to define a determinate composite acceleration then (12) is the most natural candidate definition of that point. Unfortunately there are plenty of equally natural (if not more natural) candidates such as the average position of the parts, or the position of the most central part, or the position of the most massive part, etc. None of these can be ruled out for the reasons that (15) was ruled out, yet all of these will give different results for the composite mass (given composite force and Newton's second law). Here (15) is just an easily formulated definition that helps us to see the latter point.

To account for the scientific reductive explanation of mass additivity one could either qualify the core deducibility theory of reduction to allow Feather's stipulation, or one could argue that there is a superior reduction of mass additivity that fits the core deducibility theory. In the next section I attempt to put forward such a reduction. Then in the two remaining sections I consider two deducibility theories that make distinctive qualifications to the core deducibility theory. I argue that in both cases the qualifications by themselves do not help: both theories, in their own ways, also require the explanation that I now turn to.

---

[6] E.g. let two one-dimensional particles have unit mass, and be located two meters apart at -1 and +1 respectively so that the origin is at 0. In that case (15) puts the composite at 0. But move the origin so the coordinates of the particles are now 99 and 101 and (15) puts the composite at 200√2.

**5. The Naturalness Solution to the Composite Acceleration Problem**

The central problem is that we require a determinate value for the acceleration of the composite, but when its components have distinct accelerations, composite acceleration is often *indeterminate*. By calling 'composite acceleration' indeterminate, I mean that there are several ways to assign a precise acceleration to composite objects whose parts have distinct accelerations, none of which is obviously incorrect. For a relatively complex composite whose components have (even only slightly) distinct accelerations this should be clear: there will often be no fact of the matter as to whether we should take the composite acceleration to be the average acceleration of the parts, or the average acceleration of a subset of the parts, or the acceleration of the fastest part, or the acceleration of the most massive part etc. Arguably this applies even for two-particle composites. For example if the components have distinct accelerations $a_1$ and $a_2$ in the same direction, then there may be no fact of the matter as to which of the continuum of values between $a_1$ and $a_2$ is the composite acceleration.

I will call these candidate acceleration assignments *admissible precisifications*. The approach I develop applies K&B's strategy to all admissible precisifications. Doing so yields a large number of equations that each relate a precisification to composite force and component masses. Equation (14) will be included since (13) is an admissible precisification. Meanwhile the acceleration of (15) is patently not an admissible precisification so equations involving it will be excluded. The strategy then deduces mass additivity from these equations. But is an inference from these equations to mass additivity a deduction? I believe that if David Lewis' notion of *naturalness* is part of our definitions of theoretical terms then it is. I begin by defining 'naturalness'. I then explain the strategy.

Lewis argues that the naturalness of candidate extensions plays a role in determining the extensions of our terms.[7] The naturalness of a property is defined in terms of similarity to fundamental properties (or what Lewis calls "perfectly natural properties"). For example, if more than one property is a measure of a composite's acceleration resistance then 'mass' will pick out the most natural measure (if there is one). In special relativity, composite rest mass is more like a fundamental property than composite relativistic mass because rest mass is frame-invariant. The fact that physicists now apply 'mass' exclusively to rest mass is evidence that 'mass' is applied with a naturalness constraint.

---

[7] See Lewis (1983: 370-373; 1984). Naturalness is not an external constraint on reference but is imposed by our conventions: Schwarz (2014). A formal account of the semantics of theoretical terms governed by such conventions can be given in terms of the so-called Unique Best Deserver semantic theory, see: Elliott, McQueen, and Weber (2012: sec. 4.1).

So applying 'mass' in accordance with the naturalness constraint means applying it (where possible) to the property most like a fundamental property that deserves the name. Newtonian microphysical properties are not indeterminate in any sense. So a plausible more specific naturalness constraint on 'mass' is the "best precise deserver constraint": where possible apply 'mass' to whatever *precise* value is the most accurate measure of an object's disposition to resist acceleration given applied forces.

Now consider a two-particle composite C* whose components have slightly distinct accelerations $a_1$ and $a_2$ in the same direction. The set of admissible precisifications is quite constrained. Arguably, the set is the precisifications of C*'s acceleration that are on the continuum of values between $a_1$ and $a_2$. We can (in principle) use this set together with Newton's second law, to determine which precise mass value most accurately represents the composite's disposition to resist acceleration given applied forces. The following table illustrates this using just two mass hypotheses, additive mass $m_A$ and additive mass plus (a further) unit mass $m_{A+1}$:

| Precisifications | Mass hypotheses | Second law calculations |
|---|---|---|
| $a_1$ | $m_A$ (additive mass) | $F_{C*} \sim m_A a_1$ |
| $a_2$ | | $F_{C*} \sim m_A a_2$ |
| ... | | ... |
| $a_{cm}$ | | $F_{C*} = m_A a_{cm}$ |
| ... | | ... |
| $a_N$ | | $F_{C*} \sim m_A a_N$ |
| $a_1$ | $m_{A+1}$ (near additive mass) | $F_{C*} \sim m_{A+1} a_1$ |
| $a_2$ | | $F_{C*} \sim m_{A+1} a_2$ |
| ... | | ... |
| $a_{cm}$ | | $F_{C*} \sim m_{A+1} a_{cm}$ |
| ... | | ... |
| $a_N$ | | $F_{C*} \sim m_{A+1} a_N$ |

The first mass hypothesis is that C*'s mass is additive. There is then a precisification that is *equivalent* to C*'s force divided by $m_A$: $a_{cm}$. Furthermore, because every other precisification is very close to $a_{cm}$ it follows that the product of $m_A$ and any precisification is very close to the actual force on C. So $m_A$ becomes a clear candidate for being the *most natural* measure of C*'s disposition to resist acceleration given applied forces. As we increase the mass hypothesis (e.g. $m_{A+1}$, $m_{A+2}$) or decrease the mass hypothesis (e.g. $m_{A-1}$, $m_{A-2}$), we find that the second law calculations slowly become less and less approximately true. To see this (without doing all the tedious calculations) note that as the mass hypothesis gets higher (or lower) equality eventually falls out of the third column. That is, no longer is there a precisification that is *equal* to the actual force divided by the mass hypothesis. Thus, we can rule out mass hypotheses that do not equal force divided by any precisification, since those hypotheses are inaccurate measures of acceleration resistance. And we can conclude that $m_A$ is the most accurate precise measure because multiplying it by the precisifications yields a set of values that approximate the force better than any other mass hypothesis. We can now deduce the general result that mass is additive. I offer this as a complete deductive explanation of mass additivity in Newtonian mechanics.

## 6. The Nagelian Theory of Reductive Explanation

For Nagel (1949, 1961, 1979) a reductive explanation relates a higher-level "phenomenological" theory $T_P$ to a lower-level "fundamental" theory $T_F$, where a theory is a set of statements about the world.[8] Such sets of statements are related in a reductive explanation by deduction in accordance with the core deducibility model. That is, deduction means logical deduction, where definitions can be included as premises. So for Nagel $T_P$ reduces to $T_F$ if and only if $T_P$ is logically derivable from $T_F$ together with definitions of the terms in $T_P$. The strictness of logical derivability suggests that Nagel's theory won't be able to model real reductions where $T_P$ introduces vocabulary not contained in (or definable in terms of) $T_F$. Nagel solves this problem by distinguishing between *homogeneous* reductions (where the vocabulary of $T_P$ and $T_F$ do not differ) and *inhomogeneous* or *heterogeneous* reductions (where they do), and by weakening the deductive requirements in the theory of heterogeneous reductions.

Homogeneous reductions are those in which $T_P$ and $T_F$ share a common vocabulary. In particular, "all of the "descriptive" or specific subject matter terms in the conclusion [explanandum] are either present in the premises [explanans] also or can be explicitly defined using only terms that are present" (Nagel 1979: 361). Nagel illustrates homogeneous reductions with the Newtonian explanation of (i) Galileo's law for freely falling bodies near the Earth's surface and (ii) the Keplerian laws of planetary motion (*ibid*). Heterogeneous reductions are those in which there is a term in the explanandum *not* contained in the explanans. The special qualification that Nagel introduces for heterogeneous reductions is that the explanans may be supplemented by bridge principles relating terms in the explanans with terms in the explanandum so that the deduction is possible. Standard illustrations are the reduction of the laws of thermodynamics in terms of statistical mechanics (Dizadji-Bahmani et. al. 2010: sec. 2) and the reduction of optics to electromagnetic theory (Schaffner 2012: sec. V).

The reduction of mass additivity is an example of a homogeneous reduction since 'mass' is in the explanans. However this reduction is importantly distinct from Nagel's examples (mentioned above) where Newtonian physics replaces previous theories by deducing the laws of those theories as special cases. Nagel's examples are *diachronic* reductions (Nickles 1975). These contrast with *synchronic* reductions, where $T_P$ and $T_F$ have the same (or largely overlapping) domains of application and which are simultaneously valid to various extents (Dizadji-Bahmani et. al. (2010: 393)). Thus, the reduction of mass additivity is a synchronic homogeneous reduction.[9]

---

[8] The choice of subscripts comes from Dizadji-Bahmani et. al. (2010).

A powerful objection to Nagel's theory is that real reductions as expressed by scientists are never exact deductions (Primas 1998: 83). Nagel himself (1979: 362-3) acknowledged this point in the context of diachronic homogeneous reductions and it has also been made clear in the context of synchronic heterogeneous reductions (Schaffner 2012). Contemporary Nagelians choose not to abandon Nagel's theory in the face of this. Instead, they retain the core idea that reduction works as Nagel suggested, but they introduce a new qualification. The key idea is that what is deduced from the explanans is not the exact explanandum but rather a "strong analogy" (Dizadji-Bahmani et. al 2010: 398) or "image" (Marras 2005: 342) or "cousin" (Butterfield 2011: 941) of the explanandum.

For heterogeneous reductions, one starts with $T_F$ (e.g. Newtonian microphysics) and auxiliary assumptions that simplify the system under consideration (e.g. identical component accelerations). One then deduces $T_F^*$, which is a set of statements such that replacing some of its terms using bridge principles yields $T_P^*$ where $T_P^*$ bears a strong analogy to $T_P$. Figure 1 (which is based on the figure for heterogeneous reductions in Dizadji-Bahmani et. al. 2010: 399) is intended to capture the corresponding idea for homogeneous reductions. Our mass additivity example is superimposed.

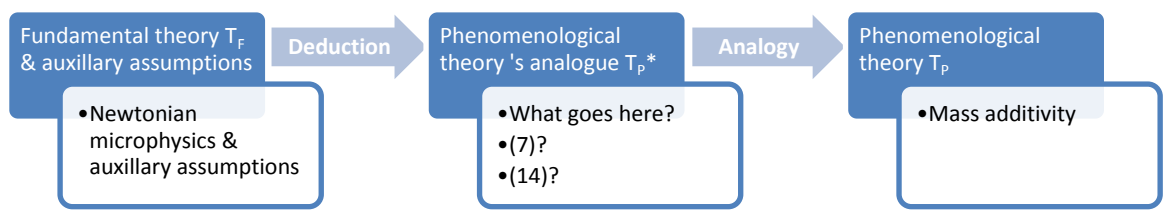

Figure 1: The Generalised Nagel-Schaffner model of synchronic homogeneous reductions. Our example of mass additivity is superimposed.

A follow-up objection to the Nagelian view is that it is unclear what it takes, in general, for a deduced theory $T_P^*$ to bear a strong analogy to the reduced theory $T_P$. Nagelians respond that other than certain weak conditions on $T_P^*$ (mentioned in a moment) there is no general story to be told, the issue must be decided on a case-by-case basis. But mass additivity poses a deeper objection. As indicated it is unclear what $T_P^*$ is supposed to be, given the textbook explanation. In fact GNS faces a dilemma: if one considers K&B's explanation then one can deduce $T_P{}^*$ (i.e. (7)) but without a strong analogy to $T_P$ (mass additivity); meanwhile if one considers Feather's explanation then there is a strong analogy between $T_P{}^*$ (i.e. (14)) and $T_P$ (mass additivity) but there is no deduction from $T_F$ to $T_P{}^*$. Either way, GNS fails to model the reduction.

---

[9] One might object that 'composite' is not contained in the reducing microphysical theory, but is clearly in the statement of mass additivity. But here one could treat unrestricted composition (or one's preferred composition principle) as a logical or a priori truth. Alternatively one could follow Dizadji-Bahmani et. al. (2010: 404) and treat the correct composition principle as an entity association law that is part of the reducing theory, as opposed to a property association law, which can't be.

If we consider K&B's explanation to be the correct explanation of mass additivity then GNS correctly models the move from $T_F$ to $T_P{}^*$. Here $T_F$ consists in Newton's microphysical laws, as well as the physical states of our particles. The crucial auxiliary assumption is the assumption of identical component accelerations. From here we deduce (7) and therefore the claim that mass is additive for composites with identical component accelerations.[10] Since this is where K&B's explanation stops, $T_P{}^*$ had better be (7). But there is no good sense in which there is a strong analogy between (7) and mass additivity. For we have been given no reason to grant that mass additivity holds in the realistic cases where component accelerations are distinct. Compare the relativistic case where mass *is* additive when components have identical velocities, but non-additive when components have distinct velocities (Okun 2009: 431). Imagine we only consider relativistic composites with identical component velocities and then deduce composite mass with relativistic laws. In this restricted case we deduce that mass is additive. But we don't thereby say that we have a strong analogy to mass additivity and that we have therefore reduced mass additivity to relativistic laws. Indeed (7) seems to clearly violate one of Dizadji-Bahmani et. al.'s (2010: 409) two conditions that (7) must satisfy if it is to bear a strong analogy to mass additivity. The first condition is that (7) use the same conceptual machinery as the principle of mass additivity. This is clear. But the second condition requires that (7) be at least as empirically adequate as mass additivity; (7) is not general enough to satisfy this constraint as the Composite Acceleration Problem shows.

But this horn of the dilemma is not inevitable. Nagelians can take comfort in the fact that K&B's explanation is controversial among some physicists, such as Feather. Perhaps Nagelians can instead treat Feather's explanation as the correct reduction of mass additivity. This would solve the strong analogy problem. After all, Feather's explanation is designed precisely so that it does not impose the restriction regarding component accelerations. So let's grant that there is a strong analogy between (14) and mass additivity. But as argued in section 4, (14) is not deduced from microphysics in Feather's explanation. Mass additivity therefore poses a prima facie dilemma for GNS.

I have posed an alternative explanation. It is therefore open to Nagelians to respond to the dilemma by treating this explanation as the correct one. In fact there are two possible ways that Nagelians can successfully model this explanation, depending on the status one gives to the inference from the set of equations involving acceleration precisifications to mass additivity. I appealed to Lewisian naturalness to argue that this inference is a deduction in the relevant sense. If it is a deduction then the appeal to strong analogy is not needed (or is trivially satisfied). On the other hand it may be controversial that this particular inference is a deduction. For example, one might object to building naturalness constraints into the definition of theoretical terms, or one might have views on indeterminacy that conflict with how the idea of a precisification has been applied. In that case, the fall-back position would be to model this part of the explanation as a rigorous defence of the strong analogy between the set of deduced equations ($T_P^*$) and mass additivity ($T_P$). Either way, the explanation offered here appears to fit the GNS model.

---

[10] Even this is not clear given the Force Additivity Problem. Thus Nagelians require my solution to the Force Additivity Problem, or something similar, to work.

## 7. A Priori Entailment and the Mind-Body Problem

There is an objection to Nagel's theory of heterogeneous reductions that is particularly influential among philosophers of mind. The objection is that it is not fine-grained enough: it cannot distinguish between the physicalist and dualist theories of consciousness.[11] Qualifying the core deducibility theory by allowing bridge principles as premises in the deduction entails (wrongly) that dualism is reductionist about consciousness. This is because dualism posits psychophysical bridge principles for the dynamical relations between consciousness and physical quantities. Dualism treats these psychophysical principles as contingent fundamental laws of nature. In opposition is physicalism which posits psychophysical bridge principles with a different status: supervenience or grounding. We need a theory of reduction that distinguishes these opposing views. So we need to be more specific about what kind of bridge principles are relevant to reduction. This is a major motivation for the a priori entailment theory of reduction (AET). In the following brief exposition of AET, all references unless otherwise stated are to Chalmers (2012).

AET is best understood in terms of more general philosophical theses called *scrutability theses* (39-60). A scrutability thesis claims that all truths are knowable from some relatively compact subset of such truths (the base truths). Scrutability theses come in different kinds and may disagree on the nature of the knowability relation and the base truths. Chalmers argues that the most philosophically significant knowability relation is *a priori entailment* and defends A Priori Scrutability, the thesis that all truths are a priori entailed by a small subset of truths (157-84).

Statement P a priori entails Q if and only if the material conditional 'If P then Q' is a priori. An important property of a priori entailment is that it may obtain in the absence of explicit finite definitions linking P and Q (12-9). Chalmers and Jackson (2001: sec. 3) illustrate with Gettier conditionals: given a sufficiently described Gettier scenario we can a priori deduce that a particular belief is not knowledge. Given a sufficiently described example of a (non-flukey) justified true belief one can a priori deduce that it is knowledge. This is possible even without a finite analytic definition of 'knowledge'. In the context of defining physicalism, Pettit (1994) illustrates the same idea with a priori inferences from 2D dot-distributions to 2D shapes. A priori conditionals of this kind are the bridge principles allowed by AET.

---

[11] See Kim (1999) and Chalmers (2012: 304-5). In what follows 'consciousness' means *phenomenal* consciousness: Chalmers (1996: 11).

For our purposes the most significant scrutability thesis is Fundamental A Priori Scrutability (404-9), which states that all truths are a priori entailed by fundamental truths, supplemented with a "totality truth" (151-6).[12] Where M is any truth, F is a complete fundamental description of reality, and T says 'and that's all that's fundamental' the key thesis states that FT→M is a priori where → is the material conditional. Fundamental truths form the reduction base for all truths. This is the background thesis that motivates AET. The idea is that Fundamental A Priori Scrutability imposes a constraint on successful reductive explanation:

> "[T]here is an important sort of reductive explanation in science for which scrutability is at least a tacit constraint. That is, it is a tacit desideratum that in principle, a given reductive story could be fleshed out with further lower-level truths, such that higher-level phenomena would be scrutable from there. [...] In practice, reductive explanations typically proceed by giving just enough detail to make it plausible that a fleshed-out story of this sort could be obtained." (307-8)

The reduction of mass additivity illustrates this. K&B take a three-particle system and from its microphysical state and Newton's laws deduce a priori that the mass of a two-particle composite is additive. Their example is suggestive of the more general a priori deduction offered here. If it were not, and if no such extension to more complex systems could be found, then arguably we should deny that mass additivity is reducible to Newtonian microphysics.

Chalmers (1996) uses AET to argue that there can be no reductive explanation of consciousness in physical terms. The strategy is to argue that all the physical aspects of reality could *conceivably* obtain in the absence of consciousness. Where P is a complete microphysical description of reality and Q says 'something is conscious', the argument is usefully summarised as follows:

(a) PT&~Q is conceivable.
(b) If PT&~Q is conceivable then PT→Q is not a priori.
(c) If PT→Q is not a priori then consciousness is not reducible to P.
(d) Consciousness is not reducible to P.

---

[12] And indexical truths (408) - a complication I ignore for present purposes.

Premise (b) is analytic since 'is conceivable' means 'cannot be ruled out a priori' (Chalmers 2002). Premise (c) is crucial to our discussion. An important defence of (c) involves case studies: successful cases of scientific reductions that are shown to be associated with a relevant a priori entailment. For example, where P stands for the microphysics of a simple Newtonian possible world and $M_A$ stands for the principle of mass additivity, we can show that the scientific reduction of $M_A$ gives enough detail to make plausible that PT→$M_A$ is a priori. And if the reduction didn't do this we would consider it to be a failed reduction. So our analysis gives indirect support to (c).

In defence of (a) Chalmers argues that the concept of consciousness cannot be (exhaustively) analysed in physically respectable (i.e. functional, structural, dynamical) terms. In other words, even if P gives enough functional, structural and dynamical information to enable us to deduce highly complex macroscopic structures, we still won't be able to deduce that there is anything it is subjectively like to be those structures. The experiential aspect of reality is left out of such stories. This is the hard problem of consciousness (Chalmers 1995).

However, our case study raises a challenge to this defence of (a). Unlike philosophers of science, philosophers of mind often involve mass additivity in their discussions of reduction e.g. Chalmers & Jackson (2001: 331). However they do not consider the scientific explanation discussed here and incorrectly treat mass additivity as an *a priori* truth. For example, according to Diaz-Leon (2011: 106) the apriority of mass additivity "seems plausible because we are using the same predicate both at the microphysical and at the macrophysical level."[13] If mass additivity is a priori then mass non-additivity is inconceivable. Now consider P: our incompletely known microphysical world. If mass non-additivity is inconceivable then P can conceivably obtain in the absence of non-additive mass. This allows us to run a parody argument for the irreducibility of non-additive composite mass. Thus, let $M_{\sim A}$ say 'something has non-additive mass':

(a') PT&~$M_{\sim A}$ is conceivable
(b') If PT&~$M_{\sim A}$ is conceivable then PT→$M_{\sim A}$ is not a priori.
(c') If PT→$M_{\sim A}$ is not a priori then non-additive mass is not reducible to P.
(d') Non-additive mass is not reducible to P.

The problem with this argument is that if P is described by special relativity then (d') is false.[14] Like premise (b), (b') is analytic. Like premise (c), (c') is supported by our discussion of Newtonian physics. So we may trace the falsity of (d') back to (a'). And we may do so no matter how strong the a priori intuition in favour of (a') is. The question is whether a physicalist can use this to argue that (d) is unwarranted because (a) is inconclusive. I will argue that the correct diagnosis of where (a') goes wrong does offer the physicalist a way out of the consciousness argument.

---

[13] See also Kim (1992: 127), McLaughlin (1997: 38), and Chalmers (2012: 291-2)
[14] To make the analogy even more vivid, let $M_{\sim A}$ instead say 'something is massive' and let P stand for a simple microphysical description of some electromagnetic radiation composed entirely of *zero-mass* photons, given in the language of special relativity. In that case (d') is shown to be false in Gabovich and Gabovich (2007).

The key point is that (a') has been asserted without full knowledge of the details in P. No matter how forceful the a priori intuition in favour of (a') seems, it is trumped by fuller knowledge of P, together with the naturalness constraint on 'mass'. To illustrate, one might initially find (a') intuitive. But by analysing the relativistic details in P one might find that there are two distinct properties that roughly fit the primary causal profile we associate with mass (i.e. the property responsible for acceleration resistance given applied forces). One is the additive 'relativistic mass'; the other is the non-additive 'rest mass'. One's a priori intuition might initially lead one to apply 'mass' exclusively to the additive property. But modern relativity theory instead privileges the non-additive property, because only rest mass is a real property of objects i.e. one that does not vary with the frame of reference (Lange 2002: 224-5). Why privilege frame-invariant properties (when it is possible to do so)? Because the universe is fundamentally frame-invariant. The relevant naturalness constraint therefore constrains the application of terms so that they apply (when possible) to invariant properties. In doing so in this case, the intuition that 'mass' cannot apply to a non-additive property is trumped.

The specific moral we can draw is that one's a priori intuition that 'mass' cannot apply to a non-additive quantity, can be (and is) trumped by the naturalness constraint on 'mass' (together with the details of P). From this we can draw a more general moral: one's a priori intuition that a given term cannot apply to a property, can be trumped by the naturalness constraint on that term (together with the details of fundamental reality). We can then apply this general moral to the consciousness argument: one's a priori intuition that 'consciousness' cannot apply to a purely structural-functional-dynamical property, can be trumped by the naturalness constraint on 'consciousness' (together with the details of P). Of course one might wonder how a naturalness constraint could possibly trump the consciousness intuition. But, prior to knowledge of relativity theory, one might equally well wonder how a naturalness constraint could possibly trump the mass additivity intuition. The key point is that without sufficient knowledge of the nature of reality, one cannot simply assert that the intuition won't get trumped. And if the simple example of mass teaches us anything, it teaches us that such intuitions are fragile in light of more detailed knowledge of reality.

Arguably, with enough relativistic information one could deduce that something has non-additive mass, and demonstrate that PT→M$_{\sim A}$ is a priori. Partial microphysical relativistic reductions of non-additive mass can be found in the physics literature.[15] Like K&B's reduction, deduction plays a significant role. These reductions are therefore more likely to support than oppose AET. The question is just whether they challenge the consciousness argument. To be clear, our considerations do not show that premise (a) is false, nor do they show that conclusion (d) is false. Rather, we've simply shown that the consciousness argument is inconclusive: we cannot claim that consciousness is not a priori entailed by a proposed reduction base merely by appeal to minimal knowledge of that base and conceptual analysis of the explanandum. The consciousness argument therefore requires a further premise: that it is unlikely that details of the reduction base will yield surprising implications about the most natural candidate referent of 'consciousness'. Since the relevant reduction base is fundamental physical reality, the required additional premise seems difficult to justify. Hence our analysis supports the theory of reduction that has been used to argue that consciousness is irreducible, but also issues a new challenge to this particular application.

---

[15] Gabovich and Gabovich (2007), Okun (2009).

## 8. Conclusion

The scientific reduction of mass additivity predominantly consists in a deduction from well understood principles of Newtonian microphysics, making it a good test case for deducibility theories of reduction. The core deducibility theory of reduction states that reductive explanation of Q in terms of P is successful if and only if Q is deducible from P (plus analytic definitions). Modern deducibility theories qualify this theory while retaining the emphasis on deduction.

The generalised Nagel-Schaffner theory (GNS) qualifies the core theory by only requiring that a strong analogy of Q be deduced from P (plus definitions). The scientific explanation of mass additivity poses a dilemma for GNS, despite the example falling into the neglected class of synchronic homogenous reductions. I have argued that to avoid this dilemma Nagelians should treat the explanation offered here as the correct explanation.

The a priori entailment theory (AET) qualifies the core theory by removing the need for finite analytic definitions. Instead, Q is reducible to P if and only if P (perhaps in conjunction with further lower-level truths) a priori entails Q. A scientific reduction is successful only if it is suggestive of such an a priori entailment. Arguably, the scientific reduction of mass additivity is successful because it is suggestive of the a priori entailment I have offered. So assuming the scientific explanation is successful it supports AET.

However, we must be sceptical of any use of AET to argue for the irreducibility of a higher-level property (like consciousness) when the lower-level truths are not fully understood. For our higher-level concepts have naturalness constraints which determine their application, and lower-level truths can yield surprising results for the most natural candidate referent for a higher-level concept. We might have a priori intuitions to the effect that non-additive composite masses cannot find a place in the microphysical world. But the microphysical world may find a place for them anyway. Similarly, we might have a priori intuitions to the effect that consciousness cannot find a place in the microphysical world. Whether or not the microphysical world does is something we can only know after we have full grasp of the microphysical world.[16]

**Appendix: A Solution to the Force Additivity Problem**

K&B successfully deduce (6) and therefore also: $F_1 = -(m_2 + m_3)a_c$. But to deduce (7) we must first deduce: $F_1 = -F_C$. Here I provide a defensible suggestion.[17]

Recall (5): $m_1 a_1 = -m_2 a_2 - m_3 a_3$. Substitution using the second law gives: $F_1 = -(F_2 + F_3)$. So we must deduce: $F_2 + F_3 = F_C$ or "the additivity of force" as that will give: $F_1 = -F_C$. I do this in four steps. Step one deduces the claim that C exerts some force on particle 1. Step two deduces that this force must be $F_1 = -(F_2 + F_3)$. Step three deduces the claim that particle 1 exerts some force on C. Step four deduces that this force must be equal but opposite to the force that C exerts on 1 i.e. $F_2 + F_3$.

Step one: necessarily, removing C from the microphysical explanans removes particle 2 or particle 3 or both. This removes force $F_2$ or $F_3$ or both. But this reduces the force on particle 1, and so given 1's constant mass and $F_i = m_i a_i$, 1's acceleration reduces. So C induces an acceleration on 1 so C exerts some force on 1.

Step two: deduce the value of $F_{1C}$ i.e. the force that C exerts on 1. We prove by reductio that $F_{1C} = F_1 = -(F_2 + F_3)$. We know that particles 2 and 3 exert $F_1 = -(F_2 + F_3)$ which (by stipulation) is the *total* force on particle 1 and we know that $F_{1C} \neq 0$ (from step one). Now assume that $F_{1C} \neq -(F_2 + F_3)$. But if $F_{1C} \neq -(F_2 + F_3)$ then the total force on 1 is not equal to $-(F_2 + F_3)$, hence a contradiction. So C exerts the force that its components exert, that is: $F_{1C} = F_1 = -(F_2 + F_3)$.

Step three: to deduce the claim that particle 1 exerts a force back on C we appeal to considerations similar to those used in step one. If we remove 1 we affect C's acceleration. Since removing (or even changing the state of) particle 1 affects C in this way there is a force on C due to 1. The remaining question is what the value of this force is.

---

[16] I would like to thank David Chalmers, Dan Marshall, Tim Maudlin, Daniel Nolan, Jesse Robertson, Raul Saucedo, Craig Savage, Wolfgang Schwarz, Jonathan Simon, Michael Simpson, Jonathan Tapsell, and two anonymous referees, for helpful feedback. This publication was made possible in part through the support of a grant from Templeton World Charity Foundation. The opinions expressed in this publication are those of the author.
[17] Prima facie to deduce $F_1 = -F_C$ just infer that C exerts $F_1$ on particle 1 in virtue of C's components exerting $F_1$ on particle 1. Then apply the third law to C and particle 1 to deduce the force on C: $-F_1$. But the third law cannot obviously be applied to more than two particles: K&B (2004: 7) and the applicability of the third law to composites is something we should deduce, not assume. So let's try another option.

Step four: to deduce the value of $F_{C1}$ we appeal to considerations similar to those used in step two. We prove by reductio that $F_{C1} = -F_1 = F_2 + F_3$. We know (by stipulation) that $F_2$ is the total force on particle 2 and that $F_3$ is the total force on particle 3, and we know (from step three) that $F_{C1} \neq 0$. Now assume that $F_{C1} \neq F_2 + F_3$. Since there can be no changes in C without changes in particles 2 and 3 then $F_{C1}$ must to some extent apply to particles 2 and 3. But then $F_2$ is not the total force on 2 or $F_3$ is not the total force on 3 (or both), hence a contradiction. So C experiences the sum of the forces that its components experience, that is: $F_{C1} = F_2 + F_3$.

Since C is only interacting with particle 1, $F_{C1}$ is the total force on C. So $F_{C1} = F_C = F_2 + F_3$ such that $F_1 = -F_C$. We now return to $F_1 = -m_2 a_2 - m_3 a_3$ and substitute: $F_c = m_1 a_1 + m_2 a_2$. Assuming $a_1 = a_2$ we then deduce (7) and solve the Force Additivity Problem. (Note that in deducing force additivity we have also deduced that the third law scales up to composites.) Thus, mass additivity is deducible from Newtonian microphysics when components have identical accelerations.